\documentclass[twocolumn,aps,prd,epsf,nofootinbib]{revtex4-1}

\usepackage{color}
\usepackage{amsmath}
\usepackage{amssymb}
\usepackage{braket}
\usepackage[T1]{fontenc}
\usepackage{url}
\usepackage[english]{babel}
\usepackage{graphicx}
\usepackage{subfigure}
\usepackage{graphicx}
\usepackage{dcolumn}
\usepackage{bm}
\usepackage{float}
\usepackage{url}

\usepackage{bclogo}

\usepackage{tikz}
\usetikzlibrary{trees}
\usetikzlibrary{decorations.pathmorphing}
\usetikzlibrary{decorations.markings}

\usepackage{amsfonts} 

\usepackage{grffile}

\usepackage{ulem}

\newcommand{\be}{\begin{equation}}
\newcommand{\ee}{\end{equation}}
\newcommand{\bea}{\begin{eqnarray}}
\newcommand{\eea}{\end{eqnarray}}

\newcommand{\bit}{\begin{itemize}}
\newcommand{\eit}{\end{itemize}}

\usepackage{verbatim}

\extrafloats{1000}

\graphicspath{{figs/}}

\begin{document}

\title{The Schwinger function, confinement and positivity violation in pure gauge QED}

\author{Lee C. Loveridge$^{1,2}$}
\email{loverilc@piercecollege.edu}
\author{Orlando Oliveira$^1$}
\email{orlando@uc.pt}
\author{Paulo J. Silva$^1$}
\email{psilva@uc.pt}

\affiliation{ \mbox{} \\ 
                 $^1$ CFisUC, Department of Physics, University of Coimbra, 3004-516 Coimbra, Portugal \\ \\
                 $^2$ Los Angeles Pierce College, 6201 Winnetka Ave.,  Woodland Hills CA 91371, USA}

\begin{abstract}
The lattice regularized pure gauge compact U(1) theory is an ideal laboratory to explore how confinement is realized as its phase diagram
has a confined and a deconfined phase that depends on the value of the coupling constant, i.e. on $\beta$.
Herein, the connection between confinement and positivity violation through the Schwinger function associated with the
Landau gauge photon propagator is investigated. 
The simulations reported show a very 
clear link between the realization of confinement and positivity violation of the photon Schwinger function and, therefore, 
of the photon K\"all\'en-Lehmann spectral density. Furthermore, a mass scale that characterizes the decay of the Schwinger function for
small time separations is computed and used to distinguish the two phases of the theory.
\end{abstract}

\maketitle



The quanta associated with the fundamental fields of QCD have never been observed as free particles.
However, in QCD, the hadron spectra and phenomenology are explained based on the dynamics of quarks and gluons.
Despite all efforts, we still do not have a non-perturbative solution for QCD and, therefore, 
the mechanism that prevents the observation of free quarks and gluons remains to be understood \cite{Alkofer:2010ue}. 
Some authors suggest that confinement of quarks is linked with particular classes of 
gluon configurations \cite{Greensite:2011zz} such as non-Abelian monopoles 
\cite{Kronfeld:1987ri,Maedan:1988yi,Kondo:1998nw,DiGiacomo:1998pm,Kondo:2018edh,Greensite:2014gra},
center vortices
\cite{Mandelstam:1974pi,Auzzi:2003fs,Cornwall:1979hz,Mandelstam:1974vf,Fujimoto:2021wsr,Eto:2021nle,Oxman:2018dzp},
etc., with some of these configurations suggesting the emergence of Abelian dominance in QCD
\cite{Nambu:1974zg,Mandelstam:1974pi,tHooft:1981bkw,Sakumichi:2015rfa}.
In general, these gluon configurations are used to give support to a (linearly) rising static potential that certainly provides an intuitive picture for 
confinement that builds on the study of non-relativistic systems.  
Explaining the quark dynamics using a potential is physically appealing, and static potentials for systems other than meson and baryons 
have been computed. 
As mentioned, these potentials are typically rising functions of the distance between constituents, for sufficiently large distances between those constituents. 
However, the description of the quark dynamics through a static potential is questionable,  specially for the lighter quarks where quantum fluctuations 
have a leading role. Despite its usefulness, the introduction of a static potential freezes the gluon dynamics and it cannot give the full picture of the QCD
dynamics.

In a quantum field theory the information on the dynamics, such as  the spectra, form factors, transport properties, etc., is summarized in its Green functions.
These functions are not necessarily easy to compute, this is particularly true outside the perturbative solution of the theory, and the extraction of 
the required information can be a complex task. Returning to the problem of the confinement mechanism in QCD, based on BRST invariance of the QCD
action, Kugo and Ojima \cite{Kugo:1979gm} suggested a confinement mechanism that strongly constrains the infrared properties of gluon and ghost propagators
the two-point correlation functions of pure Yang-Mills theory.
Interestingly, the same type of constrains were predicted using the localised Gribov-Zwanziger action for QCD
\cite{Zwanziger:1991gz,Zwanziger:1992qr} whose starting point is completely different, namely the reduction of the functional integration space.
Lattice simulations of pure Yang-Mills theory to access the gluon and ghost propagators, see e.g.
\cite{Cucchieri:2007md,Bogolubsky:2009dc,Dudal:2010tf,Duarte:2016iko} and references therein, 
did not confirm the predictions of these approaches for the propagators.
Some of these issues have been solved by extending the formulation of Gribov-Zwanziger actions; see e.g. \cite{Dudal:2007cw,Dudal:2008sp}. 
The properties of the pure Yang-Mills gluon and ghost propagators are now well established and there is a vast literature on the subject; see, for example, 
\cite{Leinweber:1998uu,Frasca:2007uz,Aguilar:2008xm,Fischer:2008uz,Boucaud:2010gr,Bornyakov:2010nc,Ilgenfritz:2010gu,Maas:2011ez,Tissier:2017fqf,Siringo:2018uho,Aguilar:2019uob,Reinosa:2020skx} 
and references therein.

The predictions of Kugo-Ojima mechanism and of the Gribov-Zwanziger action for the gluon
propagator imply the violation of positivity for the K\"all\'en-Lehmann gluon
spectral representation, that necessarily invalidate a quantum mechanical probabilistic interpretation for one-gluon particle states, meaning that
the Hilbert space of QCD does not include one-gluon particle states.
The corresponding quanta can only appear as components of the wave  function of composite states and, in this sense, they are confined 
and cannot be observed as free particles. 
In principle, that is what happens with quarks and gluons that contribute only to the internal structure of hadrons but are
at the heart of the hadronic dynamics. In this sense positivity violations imply confined quanta.
Positivity violation has been observed both for pure Yang-Mills
gluon and ghost propagators \cite{Aubin:2003ih,Fischer:2003rp,Aubin:2004av,Cucchieri:2004mf,Bowman:2007du,Dudal:2019gvn}.
In\-de\-pen\-den\-tly of the confinement mechanism it turns out that, in non-Abelian Yang-Mills theories,
positivity violation of the K\"all\'en-Lehmann spectral function is linked also with asymptotic freedom 
\cite{Cornwall:2013zra}. 
The violation of positivity condition for the spectral function associated with two-points functions has also been reported for $O(N)$ scalar theories \cite{Matsubara:1983zt} and for QCD 
at finite temperature \cite{Silva:2013rrr,Su:2014rma}.

Denoting the propagator in (Euclidean) momentum space by $D(p^2)$, its K\"all\'en-Lehmann integral representation, see e.g. \cite{Peskin:1995ev}, is given by 
\begin{equation}
  D(p^2) = \int^{+ \infty}_{\mu_0} d \mu ~ \frac{\rho( \mu ) }{p^2 + \mu} \ , 
  \label{Eq1}
\end{equation}
where the spectral function $\rho ( \mu )$ reads
\begin{equation}
   \rho ( \mu ) = \sum_{n} \, \delta ( \mu - m^2_n) \, \left| \langle 0 | \, \mathcal{O} \, | n_0 \rangle \right|^2
   \label{Eq:spectral}
\end{equation}
with $\mathcal{O}$ being the field operator associated with the corresponding quanta, $m_n$ are the masses of the physical states and
$| n_0 \rangle$ are the physical states at rest, i.e. with $\vec{p} = 0$, that are eigenstates of the four momentum operator.
It follows from the definition that $\rho (\mu)$ is positive definite. The Schwinger function
\begin{eqnarray}
C(t)  & = & \int^{+\infty}_{-\infty} \frac{d p_4}{2 \, \pi} ~  \left. \frac{}{} D(p) \right|_{\vec{p}=0} ~ e^{-\, i \, p_4 \, t}  \nonumber \\
& = & \int^{+ \infty}_{\mu_0} \frac{d \mu}{2 \, \sqrt{\mu}} ~ \rho( \mu ) ~e^{- \sqrt{\mu} \, t }
\label{Eq:Schwinger_Continnuum}
\end{eqnarray}
is given by a sum of positive terms and, therefore, is also positive definite for all $t$. From the point of view of testing confinement through positivity 
violation, the Schwinger function allows to check for the positivity of $\rho(\mu)$ without having to invert the integral equation (\ref{Eq1}).
On the other hand, if $C(t)$ is positive for all $t$, it does not provide any information on the sign for the spectral function $\rho(\mu)$. Indeed, in this case,
the contributions of the negative $\rho(\mu)$ can be cancelled by the contributions where $\rho(\mu) > 0$ resulting in a positive Schwinger function. 
In the following, we will
always assume that the propagators have an integral representation of the K\"all\'en-Lehmann type. Note that it is not clear that this assumption
holds for confined particles; see e.g. 
\cite{Binosi:2019ecz,Li:2019hyv,Falcao:2020vyr,Li:2021wol,Hayashi:2021nnj,Hayashi:2021jju}
and references therein.

Compact pure gauge QED allows us to check the connection between confinement and positivity violation.
Indeed, recently, first principles lattice simulations of the compact formulation of a pure gauge U(1) theory \cite{Wilson:1974sk}
addressed the computation of the Landau gauge photon propagator \cite{Loveridge:2021qzs,Loveridge:2021wav,Loveridge:2021zke}
\begin{equation}
   \langle A_\mu(k) \, A_\nu(p) \rangle = V \, \delta ( k + p) \, 
   \left( \delta_{\mu\nu} - \frac{p_\mu p_\nu}{p^2} \right) \, D(p^2) \ ,
\end{equation}
where $\langle \cdots \rangle$ stands for gauge average and $V$ is the lattice volume,
confirm that pure gauge compact QED has two phases. At low $\beta = 1/e^2$, where $e$ is the bare coupling constant and $\beta$ is the coupling constant that appears in the lattice version of the action,
 the theory is confining, in the
sense that the corresponding static potential grows linearly with the distance between the fermions,  while at high $\beta$ values the theory approaches
a free field theory in the thermodynamic limit. Moreover, these studies also show that in the confined phase compact QED has a mass gap and a non-trivial
topological structure, while in the deconfined phase the theory becomes massless and the topology of the gauge theory becomes trivial\footnote{The issues related to the topology
of QED and its compact formulation on the lattice are discussed in \cite{Loveridge:2021qzs,Loveridge:2021wav}.} as it should be for a 
free field theory. These properties are translated into a Landau gauge propagator $D(p^2)$ that is rather different in both phases; see e.g. Figs. 1 and 2 in
\cite{Loveridge:2021wav}. The investigation of the Landau gauge photon propagator confirms previous studies who show that the transition between the
two phases occurs at $\beta$ slightly above unity and that the transition from the confined to the deconfined phase is first order 
\cite{Polyakov:1975rs,Polyakov:1976fu,Banks:1977cc,Glimm:1977gz,Fradkin:1978th,Creutz:1979zg,Guth:1979gz,Lautrup:1980xr,Frohlich:1982gf,Berg:1983is,Kogut:1987cd,Jersak:1996mj,Panero:2005iu,Gubarev:2000eu,Arnold:2002jk,Coddington:1987yz,Nakamura:1991ww}. 

The question addressed in the current work is to explore how the connection between the transition from the confined to the deconfined phase translates into
the photon spectral function. In particular, we aim to investigate the relation between confinement and positivity violation for the case of pure gauge compact QED.

\begin{figure}[t] 
   \centering
   \includegraphics[width=3.5in]{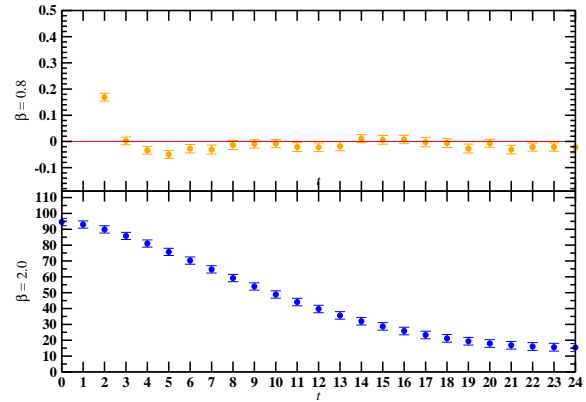}
   \caption{Bare Schwinger function for $\beta = 0.8$ and for $\beta = 1.2$ computed using a simulation on a $48^4$ lattice.}
   \label{fig:qed_schwinger_extremos}
\end{figure}

The 4D lattice simulations of the compact U(1) gauge theory considered here use hypercubic lattices with $N$ points in each direction.
For details on the sampling, gauge fixing, the  computation of the gauge fields from the links and the propagator we refer the reader to 
\cite{Loveridge:2021qzs,Loveridge:2021wav}. Note that for the computation of the Schwinger function only the so-called naive momenta are considered.
Moreover, the Schwinger function can be extracted from the Euclidean momentum propagator considering only time-like momenta. 
The lattice Schwinger function is then given by
\begin{equation}
  C(t) = \frac{1}{N} \sum^{N-1}_{n=0} ~ \left. \frac{}{} D(p)  \right|_{\vec{p} = 0} \, e^{- i \, \frac{2 \, \pi \, n}{N} \, t } \ .
  \label{Eq:LatSchwinger}
\end{equation}
For pure gauge non-Abelian theories the Schwinger function has been investigated for lattice regularized SU(2) \cite{Cucchieri:2004mf}
and SU(3)  \cite{Silva:2007sj,Silva:2006bs}  Yang-Mills theories. Further, for pure gauge theories the Schwinger function was also computed within a class of solutions 
of Dyson-Schwinger equations \cite{AlkoferPPNP03,Alkofer:2003jj}. 
In all cases it was observed that $C(t)$ takes positive and negative values, i.e. the spectral function is not positive definite.
At small $t$ the  Schwinger function decreases with time, followed by oscillations at larger $t$  often times are
around zero. 
The Schwinger function can be described by two mass scales that are associated the initial  decay and with the period of oscillation observed at larger $t$.

\begin{figure*}[t] 
   \centering
   \includegraphics[width=7in]{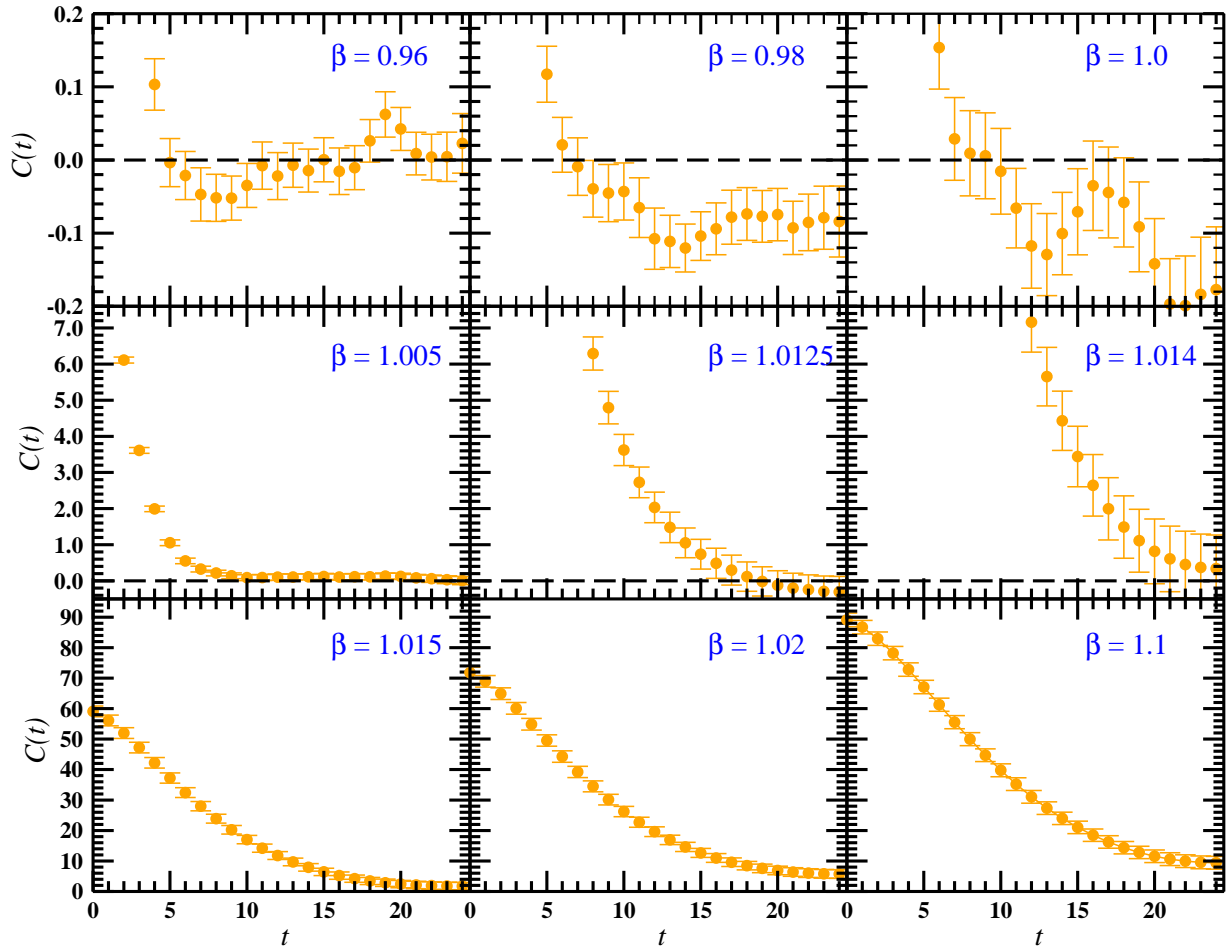}
   \caption{Bare Schwinger function close to the confinement/deconfinement transition.}
   \label{fig:qed_schwinger_various_beta}
\end{figure*}

The computation of the spectral function directly from the propagator is an ill-defined problem that is soluble after calling for regularization techniques or 
by introducing a bias. For non-Abelian gauge theories, direct attempts to measure the gluon or the ghost spectral density 
\cite{Iritani:2009mp,Dudal:2013yva,Oliveira:2016stx,
Cyrol:2018xeq,Tripolt:2018qvi,Dudal:2019gvn,Binosi:2019ecz,Li:2019hyv,Horak:2021pfr,Horak:2021syv} 
give $\rho (\mu)$ that take positive and negative values for different ranges of $\mu$ and, therefore, confirm positivity
violation in pure gauge Yang-Millg theories.

In \cite{Loveridge:2021qzs,Loveridge:2021wav}
the Landau gauge photon propagator in compact pure gauge U(1) theory was computed for various lattice volumes and $\beta$ values.
Here we calculate the Schwinger function using the outcome of the simulations reported before. 
All the quantities that are shown are bare quantities. To define a renormalized Schwinger
function one would have to renormalize the photon propagator. The difficulties of renormalizing the
two point correlation function have been discussed in \cite{Loveridge:2021wav}. 
However, the renormalization of the Schwinger function translates into a global rescale and has no impact on the observed positivity violation as a function 
of $\beta$.
The Schwinger function for the smallest $\beta = 0.8$, where the theory is in the confined phase, and largest $\beta = 1.2$, that is associated with
an essentially free theory, is reported in Fig. \ref{fig:qed_schwinger_extremos}. In the confined phase ($\beta = 0.8$), 
that shows a mass gap and a non-trivial topological structure,
$C(t)$ is positive defined for $t \lesssim 3$,
it becomes negative for the first time after $t = 3$  and then it oscillates for larger times. 
This result implies that for $\beta = 0.8$ the spectral function $\rho( \mu )$ is not a positive function and, therefore, there are no one-photon states belonging to the 
Hilbert space of the physical states in this phase. Positivity violation of $C(t)$ is not observed in the deconfined phase, where the theory is massless and topologically trivial,
and, therefore, no statement about the positivity of the spectral function can be made. The results of 
Fig. \ref{fig:qed_schwinger_extremos} suggests that for the compact formulation of QED there is a connection between confinement and positivity violation.

In order to investigate this connection deeper, the Schwinger function for various $\beta$ values computed 
around the transition between the two phases is reported in Fig. \ref{fig:qed_schwinger_various_beta}.
Positivity violation of $C(t)$ and, therefore, of $\rho ( \mu )$ is seen for $\beta < 1.005$. For larger $\beta$ the function $C(t)$ 
is always positive definite and/or becomes compatible with zero at larger $t$, within one standard deviation, and no positivity violation can be claimed.
In the confined phase, the oscillations of $C(t)$ at large $t$'s are difficult to observe but the data is still compatible with an oscillatory behaviour at larger times. 
This could be a limitation of using relatively small lattices in the simulation. Hopefully, an increase in statistics and in the lattice volume
will help identifying the oscillations with $t$. 
This oscillatory behaviour is no longer present for $\beta \geqslant 1.005$. 
For the largest $\beta$ reported, the Schwinger function is always a smooth positive function without oscillations;  
see also the bottom plot in Fig. \ref{fig:qed_schwinger_extremos}.
Furthermore, we have also checked that for $\beta = 0.8$ (confined phase) and $\beta =1.2$ (deconfined phase) 
the Schwinger functions for the largest lattice volumes considered in \cite{Loveridge:2021qzs}, i.e. the simulations using $96^4$ lattices, 
reproduce the same qualitative behaviour as observed in Figs \ref{fig:qed_schwinger_extremos} and \ref{fig:qed_schwinger_various_beta}.

\begin{figure}[t] 
   \centering
   \includegraphics[width=3.5in]{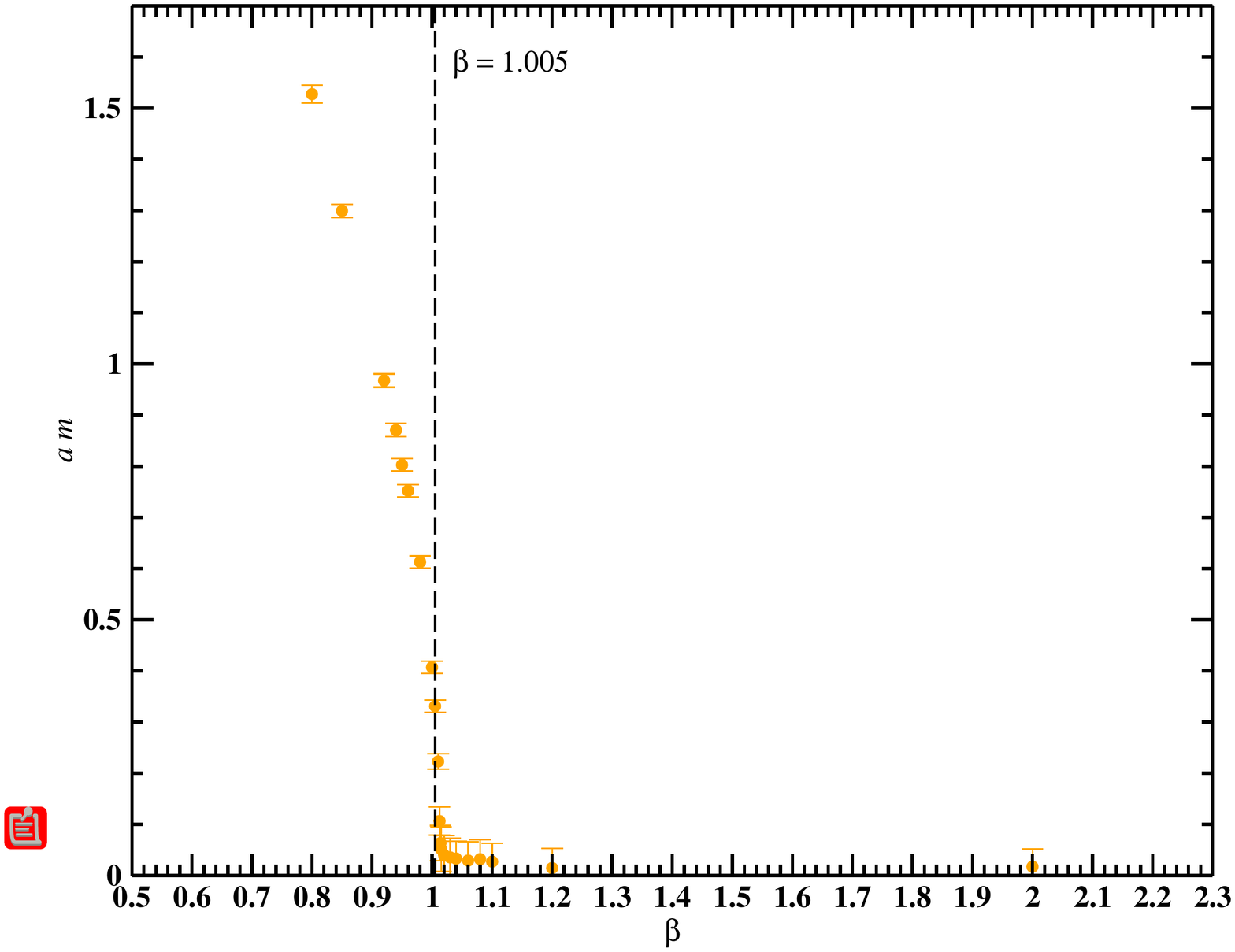}
   \caption{Effective mass associated with the decay of $C(t)$ at $t = 0$ as a function of $\beta$. The errors were computed
                  assuming Gaussian error propagation.}
   \label{fig:mass_scale}
\end{figure}

\begin{figure*}[t] 
   \centering
   \includegraphics[width=7in]{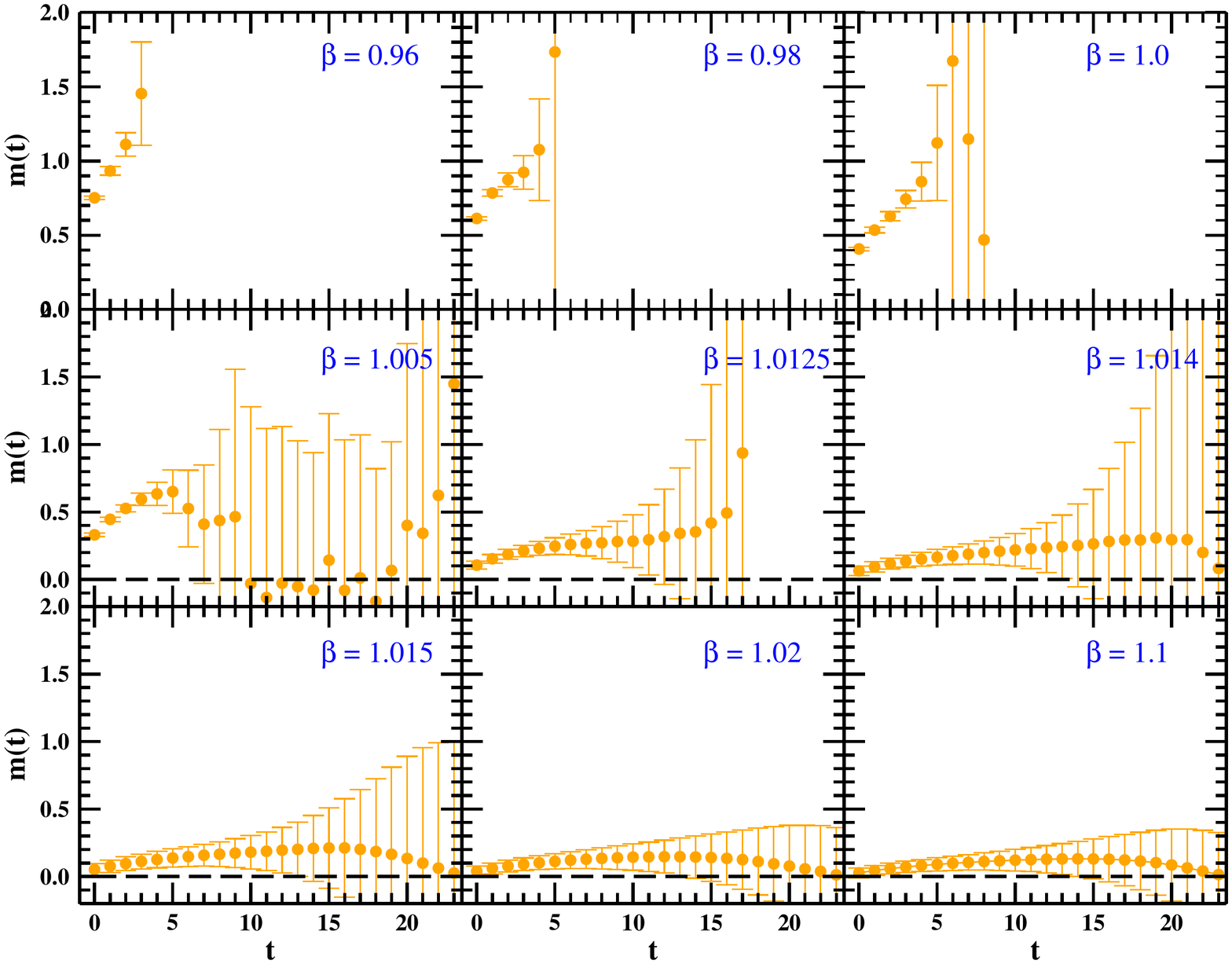}
   \caption{Effective mass $m(t)$ associated with the decay of $C(t)$ for the same $\beta$ as in Fig. \ref{fig:qed_schwinger_various_beta}. In all cases, the errors on $m(t)$ were computed
                  assuming Gaussian error propagation and the mass was defined only when $C(t)$ is positive definite.}
   \label{fig:mass_scale_t}
\end{figure*}

At small times the data for $C(t)$ decreases with $t$ both for the confined and deconfined phase. 
An effective mass scale associated with the decay can be measured taking the definition
\begin{equation}
   m(t) = - \ln \frac{C(t + 1)}{C(t)} \ .
\end{equation}
In the continuum formulation this effective mass scale is proportional to the first derivative of $C(t)$ at $t$
and it follows from Eq. (\ref{Eq:Schwinger_Continnuum}) that
\begin{equation}
m(t) \propto  \frac{1}{2} \, \int^{+ \infty}_{\mu_0} d \mu ~ \rho( \mu ) ~ e^{-\sqrt{\mu} \, t} \  .
\end{equation}
In particular $m = m(0)$ is proportional to the integral of the photon spectral density. The above mass scale
has been considered previously and studied within U(1) and SU(2) gauge theories and its variation related to positivity
violation of the spectral density \cite{Aubin:2003ih,Aubin:2004av,Cucchieri:2004mf}. Indeed, as demonstrated  in these
previous works an increase of $m(t)$ signals positivity violation. As discussed below, for compact U(1) lattice pure
gauge theory, we observe an increase of $m(t)$ at small $t$ in the confined phase and, therefore, positivity violation occurs 
in this phase. Our results for the Schwinger function confirm this result.

In all  cases that we have computed the Schwinger function, 
$C(t)$ is a decreasing function of $t$ for small $t$ and, therefore, $m$ is always positive definite for small $t$. 
$m = m(0)$ being positive definite is not in contradiction with the observed positivity violation in the confined phase. Indeed,
a $m > 0$ means that the area under the curve $y = \rho (\mu)$ for $\rho > 0$ is larger than the area under the curve 
$y = -\rho (\mu)$ where $\rho <0$. The Schwinger function is a weighted combination of these two areas and it is
the balance of the two contributions that allows or not to observe positivity violation.

In Fig. \ref{fig:mass_scale} this mass for $t = 0$, given in lattice units, is reported as a function of the coupling constant $\beta$. 
$m$ is large in the confined phase and shows a sudden drop in the vicinity of the transition towards the deconfined phase.
In the deconfined phase and for the larger $\beta$'s $m$ is compatible, within one standard deviation, with zero.
As Fig. \ref{fig:mass_scale} shows $m$ can be used to distinguish the two phases and it takes large values (in lattice units)
in the confined phase. Recall that the study of the photon propagator also shows that the confined phase has a mass gap.
Further, in  Fig. \ref{fig:mass_scale_t} the mass function $m(t)$ is given for $t$'s where the Schwinger function $C(t)$ 
has the same sign as $C(0)$, and for the same $\beta$ given in Fig. \ref{fig:qed_schwinger_various_beta}. 
Fig. \ref{fig:mass_scale_t}  complements the content of Fig. \ref{fig:mass_scale} and it also shows that, in general, $m(t)$ is an increasing function 
of $t$ for small $t$ in the confined phase, i.e. positivity violation of the spectral function occurs in the confined phase,
while in the deconfined phase $m(t)$ is essentially constant and is compatible with zero within errors. 
The study of $C(t)$ and $m(t)$ also shows clearly that, for the confined phase, even for small $t$ the Schwinger function $C(t)$ deviates from a simple exponential
decay. This can be understood from the behaviour of $m(t)$, see Fig. \ref{fig:mass_scale_t}, where $m(t)$ increases with $t$ for small times.

The reported Schwinger function  shows an oscillatory behaviour at large $t$ that appears only in the confined phase. These
oscillations define another mass that vanishes in the deconfined phase, as no oscillations are observed. We made no attempt
to access this second mass scale. Our data suggests then that the confined phase is characterized by two mass scales
that are associated with the observed decay of $C(t)$ at small $t$ and with the oscillations observed at larger times.
On the other hand for the deconfined phase, that includes the continuum limit, the theory is conformal invariant and has no 
mass scales associated with it.

In summary, the simulations reported here show, once more, that the nature of the Landau gauge photon propagator 
and also its K\"all\'en-Lehmann spectral density depend on the coupling of the compact version of the pure U(1) gauge. 
In the confined phase, the Schwinger function and, therefore, the K\"all\'en-Lehmann spectral density are not always positive definite.
On the other hand, the Schwinger function does not exhibit positivity violation in the deconfined phase,
where the continuum theory is realized. 
Our results illustrate, clearly, for the pure gauge compact QED the link between positivity violation and confinement that is expected to occur in other theories when confinement occurs.

\section*{Acknowledgments}

This work was partly supported by the FCT – Funda\c{c}\~ao para a Ci\^encia e a Tecnologia, I.P., under Projects Nos. 
UIDB/04564/2020 and UIDP/04564/2020.
P. J. S. acknowledges financial support from FCT (Portugal) under Contract No. CEECIND/00488/2017.
The authors acknowledge the Laboratory for Advanced Computing at the University of Coimbra (\url{http://www.uc.pt/lca}) 
for providing access to the HPC resource Navigator.


\end{document}